\documentstyle[psfig]{mn}
\pssilent

\def\kms{$ km \, sec^{-1}$}
\def\Mpc{\ifmmode {\, h^{-1} \, {\rm Mpc}}
\else {$h^{-1}\,$ Mpc}\fi}
\def\bfx{{\bf x}}
\def \etal {{\it et al.} }

\def\iras{{\sl IRAS}}
\def\vev#1{\langle #1 \rangle} 
\def\kms{\ifmmode {{\rm \ km \ s^{-1}}}
\else{$\rm km \ s^{-1}$}\fi}

\begin{document}

\title[The Supergalactic Plane]{The Supergalactic Plane revisited with the Optical Redshift Survey}

\author[O.~Lahav et al.]
{O. Lahav$^{1,2}$, B.X. Santiago$^{3}$, 
A.M. Webster$^{1}$, Michael A. Strauss$^{4,8}$, \and 
M. Davis$^{5}$, A. Dressler$^{6}$ \& J.P. Huchra$^{7}$ \\
$1$ Institute of Astronomy, Madingley Road, Cambridge, CB3 0HA \\ 
$2$ Racah Institute of Physics, The Hebrew University, Jerusalem 91904, Israel;
email: lahav@astro.huji.ac.il \\
$3$ Instituto de F\'\i sica, Universidade Federal do Rio Grande do Sul, 91501-970, Porto Alegre, RS, Brazil \\ 
$4$ Princeton University Observatory, Princeton, NJ 08544, USA \\
$5$ Physics and Astronomy Departments, University of California, Berkeley, CA 94720, USA \\
$6$ Observatories of the Carnegie Institution, 813 Santa Barbara St., Pasadena, 
CA 91101, USA \\
$7$ Center for Astrophysics, 60 Garden St., Cambridge, MA 02138, USA \\
$8$ Alfred P. Sloan Foundation Fellow, and Cottrell Fellow of Research
Corporation
}

\maketitle

\begin{abstract}
We re-examine the existence and extent of the planar structure in the
local galaxy density field, the so-called Supergalactic Plane (SGP).
This structure is studied here in three dimensions using both the new
Optical Redshift Survey (ORS) and the \iras\ 1.2 Jy redshift survey.
The density contrast in a slab of thickness of $20 \Mpc$ and diameter
of $80 \Mpc$ aligned with the standard de Vaucouleurs' Supergalactic
coordinates, is $\delta_{sgp} \sim 0.5$ 
for both  ORS and \iras. The
structure of the SGP is not well described by a homogeneous ellipsoid,
although it does appear to be a flattened structure, which we quantify
by calculating the moment of inertia tensor of the density field.  The
directions of the principal axes vary with radius, but the minor axis
remains within $\theta_z \sim 30^\circ$ of the standard SGP $Z$-axis,
out to a radius of $80 \Mpc$, for both ORS and \iras.  However, the
structure changes shape with radius, varying between a
flattened pancake and a dumbbell, the latter at a radius
of $\sim 50 \Mpc$, where the Great Attractor and Perseus-Pisces
superclusters dominate the distribution.  
This calls to question  
the connectivity of the `plane' beyond 
$\sim 40 \Mpc$.
The configuration found here  
can be viewed as  part of 
a web of filaments and sheets,
rather than as an isolated pancake-like structure. 
An optimal minimum variance
reconstruction of the density field using Wiener filtering which
corrects for both redshift distortion and shot noise, yields a similar
misalignment angle and behaviour of axes.  
The background-independent
statistic of axes proposed here can be best used for testing
cosmological models by comparison with $N$-body simulations.
\end{abstract}

\section{Introduction}
\label{intro}

The major planar structure  in the local universe, 
the  Supergalactic Plane (SGP), was recognized by de
Vaucouleurs (1953, 1956, 1958, 1975a,b) using the Shapley-Ames
catalogue, following an earlier analysis of radial velocities of
nearby galaxies which suggested a differential rotation of the
`metagalaxy' by Rubin (1951).  This remarkable feature in  the
distribution of nebulae was in fact  already noticed by William
Herschel more than 200 years earlier 
(for historical reviews see Flin 1986, Rubin 1989). 

When  referring to the SGP in this paper we mean the planar structure 
in the galaxy distribution. This should be distinguished 
from the formal definition of
Supergalactic coordinates. \
de Vaucouleurs \etal\ (1976, 1991) 
defined a spherical coordinate system $L,B$ in which
the equator roughly lies along the SGP (as identified at the time), 
with the North Pole ($B = 90^{\circ}$) in the
direction of Galactic coordinates ($l=47.37^\circ; b=
+6.32^\circ$). 
The position $L = 0^\circ, B = 0^\circ$ 
is at ($l=137.37^\circ, b=0^\circ$), which is
one of the two regions where the SGP is crossed by the Galactic
Plane.  The Virgo cluster is at ($L=104^\circ; B = -2^\circ$).  
Traditionally,  the Virgo cluster was regarded as the centre
of an overdense region called the `Local Supercluster'
(e.g. Davis \& Huchra 1982, Hoffman \& Salpeter 1982, Tully \& Shaya 1984, 
Lilje, Yahil \& Jones 1986).  
However, some of the much larger superclusters that are seen in recent
redshift surveys, such as the Great Attractor and  Perseus-Pisces,
are near the SGP, and  are possibly connected with
the Local Supercluster, in the sense of being simply connected by an
isodensity contour above the mean density (cf., Strauss \etal\ 1992;
Strauss \& Willick 1995; Santiago \etal\ 1995).

Although the SGP is clearly visible in whole-sky galaxy catalogues
(e.g. Lynden-Bell \& Lahav 1988; Lynden-Bell \etal\ 1988; 
Raychaudhury 1989; see also the
references above),
the extent  of planar structure 
in the galaxy distribution in 3 dimensions 
has seen little quantitative examination in recent years.
Tully (1986, 1987)
claimed that the flattened distribution of clusters extends across a
diameter of $\sim 0.1 c$ with axial ratios of 4:2:1.  Shaver \& Pierre
(1989) found that radio galaxies are more strongly concentrated to the
SGP than are optical galaxies, and that the SGP as represented by
radio galaxies extends out to redshift $z \sim 0.02$.  Stanev
\etal\ (1995) claimed that energetic cosmic rays arrive preferentially
from the
direction of the SGP, where  potential sources
(e.g. radio galaxies) are concentrated, but 
this result was criticized by Waxman, Fisher \& Piran (1997).
Di Nella \& Paturel (1995) 
revisited  the SGP using a compilation of nearly 5700 galaxies
larger than 1.6 arcmin, and found indeed that galaxies were
concentrated close to the Supergalactic Plane. 
Loan (1997) and Baleisis \etal\ (1998) searched for the SGP in projection in 
the 87GB (north) and PMN (south) radio surveys, 
and found signatures at the
1 and 3 $\sigma$ levels, respectively.

The existence of a pancake-like structure has important theoretical
implications.  Gravity amplifies deviations from sphericity, i.e.  an
initial oblate homogeneous ellipsoid evolves into a disk, and an
initial prolate structure ends up as a spindle (Lin, Mestel \& Shu
1965).  White 
\& Silk (1979) modeled the Local Supercluster as a homogeneous
ellipsoid in expanding universe, and considered implications for
cosmological models and initial conditions. Although such a simple
model is insightful, the structure of the SGP is far more complicated
than a homogeneous ellipsoid, as we show below.  In a more realistic
cosmological scenario, where the primordial density field is Gaussian,
non-spherical shapes are more abundant than spherical ones
(Doroshkevich 1970, Bardeen \etal\ 1986), and hence are expected to appear at
the present epoch as even more non-spherical.  In the context of
the `top-down' Hot Dark Matter scenario, Zel'dovich (1970) showed that
pancake-like structures are the natural outcome of gravitational
instability in the quasi-linear regime.  Further studies have
indicated that a web of filaments 
could also form  in hierarchical Cold
Dark Matter (bottom-up) scenarios (e.g. Bond, Kofman \&
Pogosyan 1996), but the shapes look quite different in different
scenarios, as a result of both initial conditions and the cosmic time
which allows the perturbations to grow.  Hence, 
quantitative measures
of the SGP and other observed filamentary structures and superclusters
(Bahcall 1988) could be very important in distinguishing between
models, e.g.  by applying the shape statistic to both data and to
$N$-body simulations.

Here we study the properties of the SGP using the Optical
Redshift Survey (ORS, Santiago \etal\ 1995) which provides the most
detailed and uniform optically-selected sample of the local galaxy
density field to date.  For comparison, we also use the full-sky
redshift survey of galaxies selected from the database of the {\it
Infrared Astronomical Satellite} (\iras), complete to 1.2 Jy at
60$\mu$m (Fisher \etal\ 1995a).  In this paper we consider the
approach of `moment of inertia' (MoI) to quantify a planar structure
(cf. Babul \& Starkman 1992; Luo \& Vishniac 1995; Dave \etal\ 1997; 
Sathyaprakash \etal\ 1998).

The outline of the paper is as follows. Section 2 describes the ORS
and \iras~ samples, while in Section 3 we discuss the visual impression and
preliminary analysis of the SGP.  Section 4 describes the formalism of
the MoI we also develop related statistics which are independent of
the background determination.  Section 5 gives the
results of MoI when applied to the samples, and Section 6 presents
Wiener reconstruction of the MoI for \iras.  Section 7 gives
interpretation of the results using mock realizations, and conclusions
and future work are discussed in Section 8.

\section{The ORS and \iras\  samples}
\label{samples}

The Optical Redshift Survey (ORS, Santiago \etal\ 1995)
covers the sky at Galactic latitude $|b|>20^\circ$.
The survey was drawn from the 
UGC (Nilson 1973), ESO (Lauberts 1982), and ESGC (Corwin \& Skiff
1995) galaxy 
catalogues, and it contains two subsets: one complete to a blue 
photographic magnitude of 14.5, and the other complete to 
a blue major diameter of 1.9 arcmin.
The entire sample consists of 8457 galaxies, with redshifts 
available for 98\% of them; $\approx 1300$ 
new redshifts were measured to complete the survey.
The high number density of galaxies 
in ORS makes it ideal 
for cosmographical studies of the local universe.

As ORS only covers $|b|> 20^\circ$, 
we filled in the Zone of Avoidance (ZOA) 
at $|b|<20^\circ$
with galaxies from the \iras\  1.2 Jy survey (Fisher \etal\ 1995a).
The Zone of Avoidance in \iras, 
$|b|<5^\circ$, 
was filled  by interpolation   
based on the observed galaxy distribution below and above the ZOA
(Yahil \etal\ 1991).
While in principle it is possible to interpolate for the ORS ZOA, 
e.g. by hand or using a Wiener reconstruction (Lahav \etal\ 1994),
we prefer to include real structure as probed by 
\iras\  even at the price of mixing two different data sets.
Hereafter when we refer to the ORS sample we mean 
that 
supplemented by the \iras\ 1.2 Jy galaxies at $|b|<20^\circ$.

Due to Galactic extinction and the diversity of catalogues used,
the selection function of ORS depends on both distance and direction
(Santiago \etal\ 1996).
In order to account for these selection effects, a weight is 
usually associated with each galaxy.
For a uniform survey like \iras\ (see below)
the selection function $\phi$ depends only on the distance to a 
galaxy,  $|{\bf x}|$,
and hence the weight is simply 
$$
w_{gal} = { 1 \over { \vev{n} \phi(|{\bf x}|)} },
\eqno (1) 
$$
where the mean number density of galaxies 
is estimated by 
$$  
 \vev{n} =  { 1 \over V} \sum_{gal} 1/\phi(|{\bf x}|)\;
\eqno (2)
$$
(see Davis \& Huchra 1982 for an alternative 
minimum variance estimator of the mean density). 

For ORS, we follow Santiago \etal\ (1996) and Hermit \etal\ (1996) and
weight each galaxy by:
$$
w_{gal} =  { 1 \over { \vev{n}_\mu \phi({\bf x}) }   } \; 
{ \vev{n}_{\mu,iras}  \over \vev{n}_{tot,iras} }\; ,
\eqno (3)
$$
where $\mu$ labels the catalogue from which the galaxy in question is
drawn; each of the four catalogues making up the ORS has its own mean
density $\vev{n}_{\mu}$. Here  $\vev{n}_{\mu,iras}$ is the mean number density 
of \iras\ galaxies in the volume corresponding to the $\mu$-th
catalogue in ORS.  This brings each of the subsamples, with their
separate selection criteria, to a uniform weighting (Santiago \etal\
1996). 
With the weighting of equation (1) or (3), we can thus calculate the
galaxy density {\it fluctuation field} $\delta(\bfx) \equiv
(\rho(\bfx) - \vev{\rho})/\vev{\rho}$ throughout the volume surveyed. 

For comparison we also apply our analysis to  the complete \iras\  1.2
Jy sample
(Fisher \etal\ 1995a; Strauss \etal\ 1992), which includes 5313 
galaxies over 87.6 \% of the sky. 
The ZOA in \iras\  was interpolated according to the Yahil \etal\ (1991)
procedure mentioned above.
Hereafter we refer to this sample as 
the \iras\  sample. 
We note that the ORS and \iras\ samples are not really independent;
roughly 60\% of the \iras\ 1.2 Jy galaxies out to $80 \Mpc$ 
are also catalogued in the 
ORS magnitude limited sample. 

We work purely in Local Group redshift space, except in section 6 on 
Wiener reconstruction.

\section{ A BRIEF TOUR OF THE SGP}
\label{quicklook}

\subsection{Visual impression}

\begin{figure}
\begin{center}
\parbox{17cm}{\psfig{figure=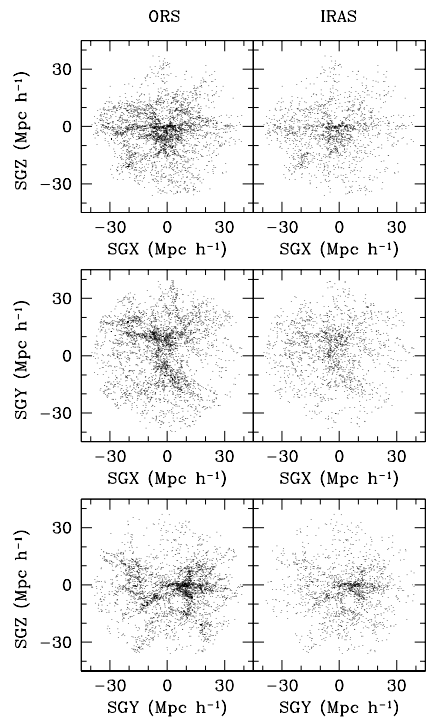,width=10cm,clip=}}
\end{center}
\caption{\label{sgp_xyz} 
The distributions  of ORS and \iras\  galaxies
in a sphere of radius of $40 \Mpc$, 
projected on the standard ($SGX$-$SGZ$), ($SGX$-$SGY$) and ($SGY$-$SGZ$)
planes. The SGP is visible edge-on as the linear feature at $SGZ = 0$
in the ($SGX$-$SGZ$)
and ($SGY$-$SGZ$) projections.}
\end{figure}

\begin{figure}
\begin{center}
\parbox{12cm}{\psfig{figure=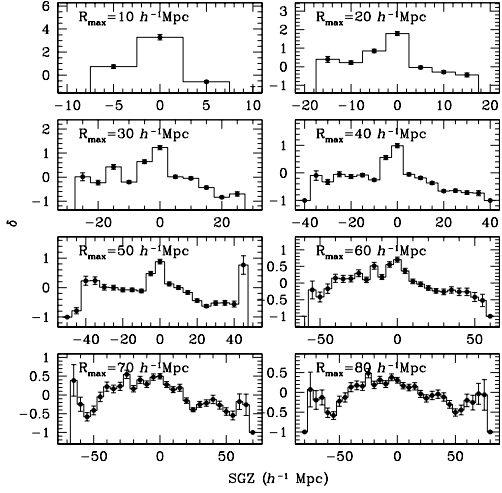,height=7cm,width=7cm,clip=}}
\end{center}
\caption{
\label{ors_sgz}
Histograms of the ORS density
field (corrected for selection effects)
as a function of $SGZ$, averaged over $SGX$ and $SGY$, 
within spheres of radius as indicated 
in each panel. The error bars are Poissonian. }
\end{figure}

\begin{figure}
\begin{center}
\parbox{12cm}{\psfig{figure=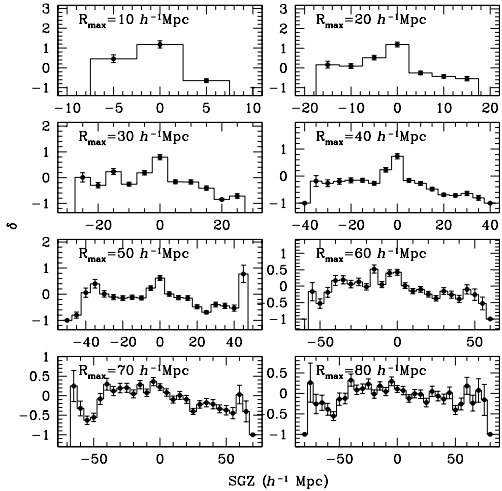,height=7cm,width=7cm,clip=}}
\end{center}
\caption{
\label{iras_sgz} 
Histograms of the \iras\ density
field (corrected for selection effects)
as a function of $SGZ$, averaged over $SGX$ and $SGY$, 
within spheres of radius as indicated 
in each panel. The error bars are Poissonian.}
\end{figure}

Historically, the SGP was identified in projected 2-D maps.  We used
the \iras\ sample (which is homogeneous over the sky) to revisit the 2-D
appearance of the SGP by fitting the data to a great circle.
For a projected distribution of sources 
the great circle along which 
the density of galaxies is enhanced can be found by 
calculating the covariance matrix:
$
I_{ij} = \sum_{gal}  {\hat x_i} {\hat x_j} 
$
where the Cartesian axes  ${\hat x_i}$ ($i=1,2,3$) are defined 
for a unit sphere. 
The matrix $I_{ij}$ can be  diagonalised (cf. Section 4.2) and the 
smallest eigen-vector indicates the direction of the normal to 
the great circle.
We considered the projected \iras\ galaxy distribution
out to $40 \Mpc$ and $80 \Mpc$, and found that the great circle is
aligned with the standard de Vaucouleurs' SGP to within $19^\circ$ and
$7^\circ$ respectively.  Can we see the translation of the great
circle to a plane in 3-dimensions?  Figure~\ref{sgp_xyz} shows the
raw distributions, uncorrected for selection effects of ORS and \iras\
galaxies in a sphere of radius of $40 \Mpc$ projected on the standard
($SGX$-$SGZ$), ($SGX$-$SGY$) and ($SGY$-$SGZ$) planes. The SGP is particularly
visible edge-on in the ($SGX$-$SGZ$) and ($SGY$-$SGZ$) projections.  We note
that the structures seen in ORS and \iras\ are quite similar, but the
ORS map is much denser, and clusters are more prominent.
Maps of the smoothed density field, after applying the appropriate
weights, 
are shown elsewhere for the \iras\ survey  (Saunders \etal\ 1991;
Strauss \etal\ 1992; Fisher \etal\ 1995b;  Strauss \& Willick 1995;   
Webster, Lahav \& Fisher 1997) and for ORS 
(Santiago \etal\ 1995, 1996; Baker \etal\ 1998). 
In Figures~\ref{ors_sgz} and~\ref{iras_sgz} we show 
histograms of the galaxy density field $\delta$, corrected for
selection  effects, 
as a function of $SGZ$, averaged over $SGX$ and $SGY$, 
within spheres of ever-larger radii, as indicated 
in each panel.
The SGP is seen as an enhancement at $SGZ = 0$ in these plots to
roughly $R_{max} = 60 \Mpc$, but is not apparent on larger scales.

\subsection {The level of overdensity} 
Our visual impression is that there is indeed a flattened structure
in the galaxy distribution, aligned along $SGZ = 0$, extending to
appreciable redshifts. 
To  quantify this 
we calculate the overdensity 
$\delta_{sgp} \equiv  n_{sgp}/\vev{n} - 1$ as a function of 
$R_{max}$, 
where $n_{sgp}$ is evaluated within 
a  circular
cylinder centered on the Local Group with radius $R_{max}$ and height
along SGZ of $20 \Mpc$, 
and the mean density $\vev{n}$ is calculated within a sphere of radius
$R_{max}$ ({\it including} the slab).
This overdensity reaches a maximum at $R_{max}=40 \Mpc$
(for which the volume of the slab is $\sim (46 \Mpc)^3$),
$\delta_{sgp}(40) =0.48 \pm 0.06 $ and $0.46 \pm 0.09$ 
for ORS and \iras, respectively.
At $R_{max}=60 \Mpc$,
$\delta_{sgp}(60) =  0.36 \pm 0.06$ and $0.26 \pm 0.08$  
for ORS and \iras, respectively.
For reference, the rms fluctuations in cubes of volume $(40 \Mpc)^3$
are $\delta_{rms} \sim 0.37 $  and $\sim0.30$ for optical (Stromlo/APM)
and \iras\ (1.2 Jy) surveys, respectively (Efstathiou 1993). 
Hence the fluctuation in galaxy counts 
in a slab  aligned with the standard de Vaucouleurs' SGP is no more than 
$\sim 1.5\,\sigma$ perturbation
on scale of $\sim 40 \Mpc$.  This implies that the SGP is only 
a modest perturbation on these large scales.
The overdensity as a function of the $Z$ component of the standard SGP 
is depicted in Figures 2 and 3.

\subsection{The centre of the overdensity}

\begin{figure}
\begin{center}
\parbox{10cm}{\psfig{figure=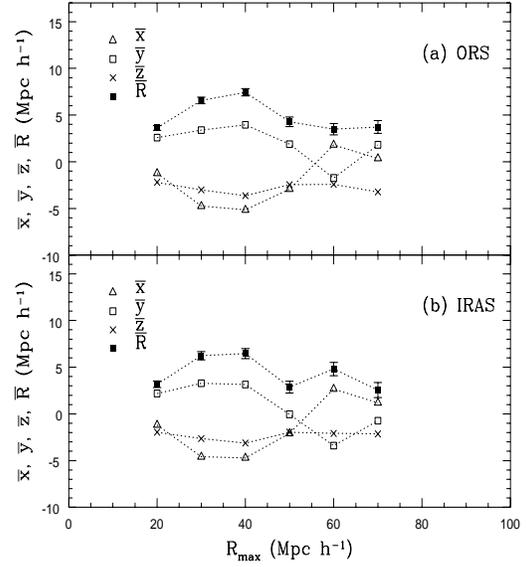,height=8cm,width=7cm,clip=}}
\end{center}
\caption{\label{distc} 
The variation of the mean position
of the centre of mass of the galaxy distribution
within a sphere of radius $R_{max}$ centred on
us, for both the ORS and \iras\
samples.
The curves show the Supergalactic $x,y,z$ of the centre of mass,
and its radial distance from the origin.  
}
\end{figure}

In order to quantify the local overdensity, we need to calculate the
centre of `mass' of the density field, 
within a sphere of radius $R_{max}$ centred on 
us, for both the ORS and \iras\ 
samples (cf., Juszkiewicz \etal\ 1990).
The centre of mass is given by summing over the galaxies (assuming they 
all have equal mass):
$$
{\bar x_i} = {1 \over  V }  
\sum_{gal} w_{gal} \; x_i \;, 
\eqno (4) 
$$
where $i=1,2,3$, $V$ is the total volume of the sphere
of radius $R_{max}$ and $w_{gal}$ is the weight per galaxy 
described in \S 2. 
The shot noise error in each axis is 
$$
\sigma_{\bar x} ^2 = {1 \over V^2} \sum_{gal} w_{gal}^2 \;x_i^2
= { 1 \over 3 V^2} \sum_{gal} w_{gal}^2 \;r^2,
\eqno (5) 
$$ 
where the last equality is valid for an isotropic distribution.
Figure~\ref{distc} shows the coordinates of the centre of the galaxy
distribution as a function of $R_{max}$.  The centre 
moves away from the origin (the Local Group) 
by no more than $8 \Mpc$  for ORS and $6 \Mpc$  for \iras. 
This is partially due to the `tug of war' between the 
Great Attractor and Perseus-Pisces, which largely balance each other
out. Indeed, the centre
moves back towards the origin for $R_{max} > 40 \Mpc$.

\section{Principal Components of the Inertia Tensor}
Both visual impression of 
and a formal $\chi^2$ fit suggest that the SGP 
cannot  be fitted by 
a homogeneous ellipsoid model. 
We therefore seek a more objective measure of deviation from 
sphericity, via the Moment of Inertia tensor (MoI).

\subsection{Estimation of the Moment of Inertia}

We wish to detect  a 
high density 
planar structure (e.g. a slab or an ellipsoid) embedded in a 
uniform sphere of radius $R_{max}$. 
One approach 
(cf. Babul \& Starkman 1992, Luo \& Vishniac 1995, Dave \etal\ 1997) 
is to construct the 
MoI 
for the {\it fluctuation} in the density field:
$$
{\tilde  C_{ij} }  = \; C_{ij} - {\bar {C_{ij}} } = \;   
 {1 \over { \vev{n} V} }  \int  [n({\bf x}) -  n_{bg} ] 
(x_i - {\bar x_i }) (x_j - {\bar x_j}) \; dV, 
\eqno (6)
$$ 
where 
$n({\bf x})$ is the galaxy number density at position ${\bf x}$,
$\vev{n}$ is the mean number density 
and $x_i, x_j$ ($i,j=1,2,3$) are Cartesian components of ${\bf x}$. 
Note that 
we allow the centre $\bar {\bf x}$ to move (see Figure~4), although
below, we will end up keeping it fixed.
The integration is over the volume of the sphere.
We define the fluctuations in the density field relative to 
the {\it background} density $n_{bg}$ in the absence of the slab
(which differs from the mean density $\vev{n}$ {\it including} the slab).
 
For a uniform distribution with density $\vev{n}$
the covariance matrix is analytic and isotropic:
$$
{\bar C_{ij} } = { f_{bg} \over  V } \int  x_i x_j dV = 
{1 \over 3} \delta_{ij}^K {f_{bg} \over V }
\int_0^{R_{max}} r^2 (4 \pi r^2 dr) =
$$

$$
=\delta_{ij}^K \; f_{bg}\;
{ R_{max}^2 \over 5 },
\eqno (7) 
$$ 
where 
$$
f_{bg} \equiv  {n_{bg} \over {\vev{n}} }  ,
\eqno(8)
$$
$\bar {\bf x} $ is of course zero, and  $\delta_{ij}^K$ is the Kronecker
delta function. 

For a  discrete density field, such as provided by ORS or \iras\ ,
we can calculate the covariance elements
by summing over the galaxies:

$$
{\tilde C_{ij} }  = C_{ij} - {\bar C_{ij} } = 
{1 \over   V }  \sum_{gal} w_{gal} 
x_i x_j  -
\delta_{ij}^K f_{bg}  { R_{max}^2 \over 5 } \;,
\eqno (9)
$$
where we have set $\bar {\bf x} = 0$, given the fact that the centre
of mass stays so close to the origin (Figure~4). 
An analytic estimator for Poisson shot-noise per diagonal component
of the covariance matrix (assuming no errors in $f_{bg}$) is: 
$$
\sigma_c^2 = { 1 \over {(3 V)}^2  } 
\sum_{gal}  w_{gal}^2 \;  r^4 \;. 
\eqno (10) 
$$  

Eq. (9) requires knowledge of $f_{bg}$ 
in order to correct for the background density.
However, as the boundaries of the SGP are hard to define {\it a priori}
and the background will not be uniform,
it is difficult to estimate a meaningful $f_{bg}$. 
In \S 4.3 we suggest a statistic which is independent of this parameter.

\subsection{Principal axes}

The next step in our analysis is to diagonalize the MoI and
find the eigenvalues $\lambda_\alpha$ and associated eigenvectors ${\bf
u}_\alpha$ ($\alpha=1,2,3$): 
$$
{\tilde C}  {\bf u}_\alpha = \lambda_\alpha {\bf u}_\alpha \;. 
\eqno (11)
$$
The standard deviation (`1-$\sigma$') along each of the three axes is given by 
$\sqrt{\lambda_{\alpha}}$, which we label hereafter as $a,b,c$. 
Note that since the background contribution 
(the last term in eq.~9)
is isotropic, it only affects the
eigenvalues, 
but not the directions of the 
eigenvectors. 
This procedure is essentially the 
Principal Component Analysis (PCA) -  a well known statistical 
tool for reducing
the dimensionality of parameter space
(e.g. Murtagh \& Heck 1987 and references therein).
By identifying the {\it linear}
combination of input parameters with maximum variance, PCA 
determines  the Principal Components
that can be most effectively used 
to characterize the input data. 
In our case, in searching for a plane, we wish 
to see if the PCA finds one axis to be much shorter 
than the other two axes. 


\subsection{Background-independent statistic of axes}

Our aim is to quantify deviations from spherical structure.
The difficulty  in doing so is due to two problems:
(i) the background density as modeled above with $f_{bg}$
is ill-determined, 
and (ii) the shot noise due to the finite number of galaxies can be
quite large. 

We will discuss the shot noise problem in \S 6 below.  To overcome the
problem of the unknown background, 
we can construct the following quantities which are independent 
of  $f_{bg}$:
$$
p^2 = |a^2 -b^2| \;, 
\eqno (12)
$$

$$
q^2 = |a^2 -c^2| \;.
\eqno (13) 
$$
and
$$
s^2 = |b^2 -c^2| \;.
\eqno (14) 
$$
For a perfect sphere, $p=q=s=0$.
For a homogeneous oblate ellipsoid with semi-major axes 
$A = B >  C$, one expects 
$p=0$ for all $R_{max}$, but for $q$ and $s$ 
to increase out to  $R_{max} \approx A$ and then to decline.
Note that the $(p,q,s)$ 
here are in the diagonalized PCA frame, 
where  the  axes  are ordered by size ($a \ge b \ge c$).
Alternatively, we can calculate $(p,q,s)$
along fixed 
$(x,y,z)$ axes (e.g. in de Vaucouleurs' system) by replacing $a^2,b^2,c^2$ 
in eqs. (12-14) by 
the non-ordered variances in that fixed coordinate system 
(i.e. the diagonal elements in the MoI matrix in that $(x,y,z)$ coordinate system).

In the limit that the errors
are isotropic the Poisson error bars are: 
$$
\Delta p = \sigma_c^2/p \;, 
\eqno (15)
$$
and similarly for $q$ and $s$.
We will determine the characteristic shape of the galaxy density
field, from the observed  $p,q$ and $s$ and their error bars
as a function of $R_{max}$.
We note that  other transformations of $(a,b,c)$ are possible,
such as the $(S1,S2,S3)$ statistic proposed by Babul \& Starkman (1992).

\subsection{Mock realizations} 

\begin{figure}
\begin{center}
\parbox{10cm}{\psfig{figure=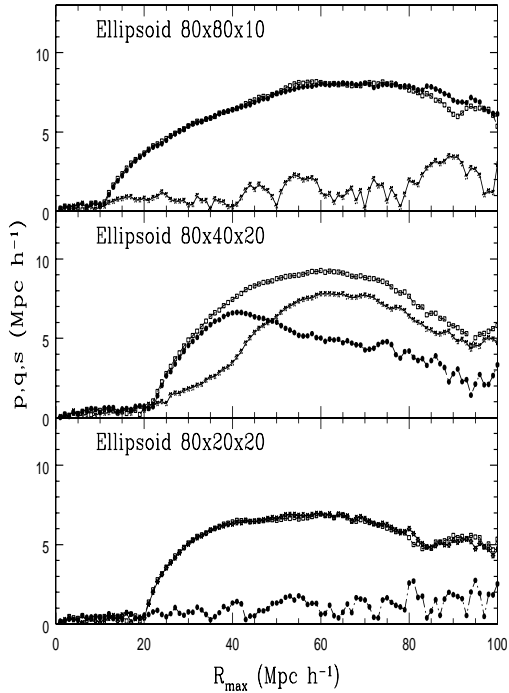,height=10cm,width=7cm,clip=}}
\end{center}
\caption{\label{elips} 
 The quantities $p$ (crosses), $q$ (open squares) 
and $s$ (filled circles)  given by eqs.~(12-14) vs.~$R_{max}$  for
             mock ellipsoids (oblate, prolate and triaxial) with 
             dimensions (in $\Mpc$) indicated in the panels, 
             all ellipsoids with density contrast of $\delta = 0.4$.
             The mock samples have an \iras\ selection function,
             and the results are averaged over 100 realizations at
fixed axes.  
}
\end{figure}

To get an insight into our $(p,q,s)$ statistic we applied it to mock
realizations of ellipsoids placed in a uniform background.  The \iras\
selection function was applied 
to the mock samples, and the results were averaged over 100
realizations.  Figure~\ref{elips} shows the quantities $p, q$, and
$s$ derived at fixed axes for mock ellipsoids (oblate, prolate and
triaxial) with dimensions indicated in the panels and overdensity of
$\delta=0.4$.  When the axes were allowed to be chosen by PCA,
the axes agreed with the correct orientation of the ellipsoids to within
$7^\circ$ (rms over 100 realizations).  
For
a single realization the curves are much more noisy.  Moreover, if 
PCA defines the axes, it attempts to
maximize the differences between the axes, and the results for $p, q$,
and $s$ are biased by noise.  For example, in the case
of a perfectly oblate structure in the presence of noise, the quantity
$s$ increases with radius $R_{max}$, where it is identically zero in
the noise-free case. 
Fortunately, we will see in the following section that the behaviour
of $(p,q,s)$ for the real 
data is quite robust. We will continue these experiments with the mock
realizations in \S 7, where we consider more complex geometries.

\section {Application of MoI to ORS and IRAS}

   We now apply the MoI approach to the ORS and \iras\ samples.
   We begin by considering $p,q,$ and $s$ along {\it fixed}  axes.  We
will find below that the PCA direction at small $R_{max}$ is indeed
very close to that of de Vaucouleurs' system, so we simply choose  the
standard de Vaucouleurs' SGP axes.  
Figure 6 shows $p,q$ and $s$ 
   (derived here from the variances along the 
   standard SGP axes without ordering them, as explained in section 4.3)
   as a function of $R_{max}$.  
  The most notable feature is the dip in
   $s$ at about $R_{max} =50 \Mpc$ in both samples. 
   This is  due to the dumbbell-like 
   configuration
   of Perseus-Pisces and the Great Attractor on opposite sides of the
   sky, as we'll see in \S 7.  But 
   at $R_{max}=80 \Mpc$, the structure is pancake-like: $p$ tends
   to small values 
   (in particular in ORS), more in line with the 
   visual impression
   of a pancake-like structure in Figure~1.
   However, the dramatic changes with $R_{max}$ (in particular 
   at $50 \Mpc$)  call to question the 
   notion of a single coherent feature within the boundaries of our
   samples (see \S 7).  
   Based on Figure 6, one may argue that the `plane' terminates at 
   $\sim 40 \Mpc$.

    We now turn to the  PCA approach of diagonalizing the 
   MoI derived at each radius $R_{max}$. 
   In this case the axes change direction, and might even `swap'. 
   'Swapping' means  that an axis which is  the largest at a given
   $R_{max}$ may become say the second or third largest in a different
   $R_{max}$. Hence we may see a discontinuity in the variation of the 
   `largest' axis with $R_{max}$. 
  
   The variation at large $R_{max}$ is more difficult to 
   interpret, but we see that $q>p>s$, implying a triaxial shape, 
   and $p$ and  $q$  increase with $R_{max}$.
   In contrast, the toy models (Figure 5) show 
   that two of $[p,q,s]$  drop with $R_{max}$ beyond the radius which encloses 
   the ellipsoid ($80 \Mpc$ in the toy models). 
   But in making this comparison with Figure 5
   we should keep in mind  that at large $R_{max}$ the increase of the 
   observed quantities $p,q$ or $s$ might be due to shot-noise.
   It is also interesting to note  the similarity between ORS and \iras, 
   indicating that on the very large scales the density fields in the two
   samples  are quite similar.

   Table 1 lists $a,b,c$, their Poisson errors, and the direction of
   the PCA  $Z$ component  
   for ORS (for direct summation) and \iras\ (for both direct summation and 
   Wiener reconstruction described in the next Section). 
   Table 1 also lists the values in {\it differential} shells, 
   showing  trends similar to those seen in the cumulative plots.

   Fig 9 shows the variation of direction of the PCA $Z$ component
with $R_{max}$.
We see that although 
the angle 
$\theta_z \equiv 90^\circ - B_z$ 
between the PCA minor axis and  de Vaucouleurs' pole
is usually less than $30^\circ$, 
there is a significant  variation on the sky
(relative to the errors of $\sim 10^\circ$ due to shot noise).  
Particularly noticeable is the  
big jump at $R_{max} = 50 \Mpc$, in accord with the 
behaviour in  Figures 6,7 and 8. 
This is   again due to the 
dumbbell configuration of  the Great Attractor 
and the Perseus-Pisces superclusters, as we shall see in \S 7.
Table 1 also gives $B_z$, the complement of the misalignment angle
$\theta_z$, for 3 shells, each  20 \Mpc\ wide.
We see that indeed  typically $\theta_z \sim 30^\circ$, 
even at the (60-80) \Mpc\ shell, which is beyond the Great Attractor 
and Perseus-Pisces.
The probability that two random vectors 
are separated by  an angle less than  $\theta_z \sim 30^\circ$
by chance
is $P(<  \theta_z) = [1-\cos(\theta_z)]$, i.e. $\sim 13 \%$.

\begin{figure}
\begin{center}
\parbox{10cm}{\psfig{figure=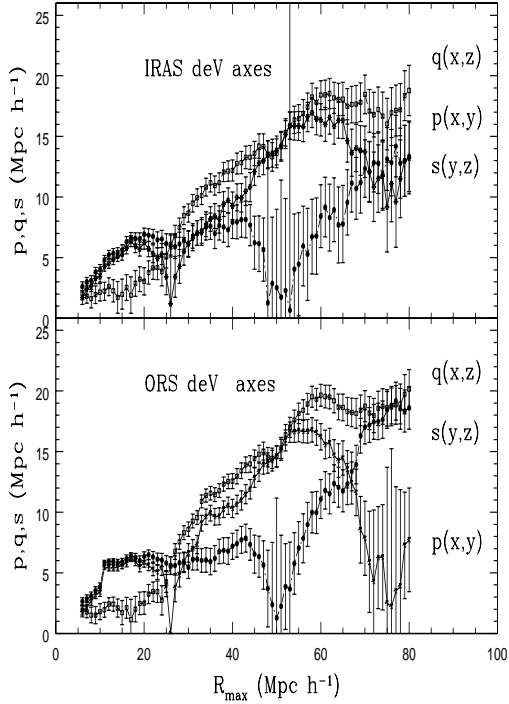,height=10cm,width=7cm,clip=}}
\end{center}
\caption{\label{pqs_dev} 
 The quantities  $(p,q,s)$ derived from the variances 
 along  the de Vaucouleurs' axes 
 vs. $R_{max}$
 (cumulative) 
 for the ORS and \iras\ samples.  The error
 bars are Poisson.}
\end{figure}

\begin{figure}
\begin{center}
\parbox{10cm}{\psfig{figure=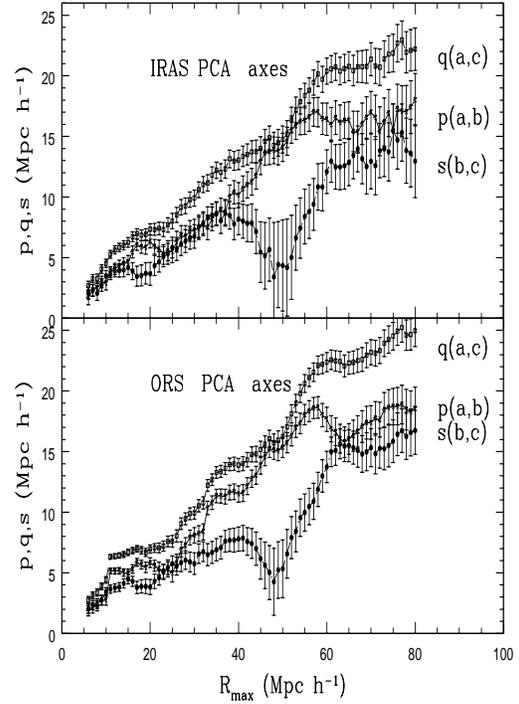,height=10cm,width=7cm,clip=}}
\end{center}
\caption{\label{pqs_pca} 
The quantities $(p,q,s)$ derived from  PCA dimensions 
$(a,b,c)$ (eqs. 12-14)
vs. $R_{max}$ (cumulative) 
for the ORS and \iras\ samples.  The error
bars are Poisson.}
\end{figure}

\begin{figure}
\begin{center}
\parbox{15cm}{\psfig{figure=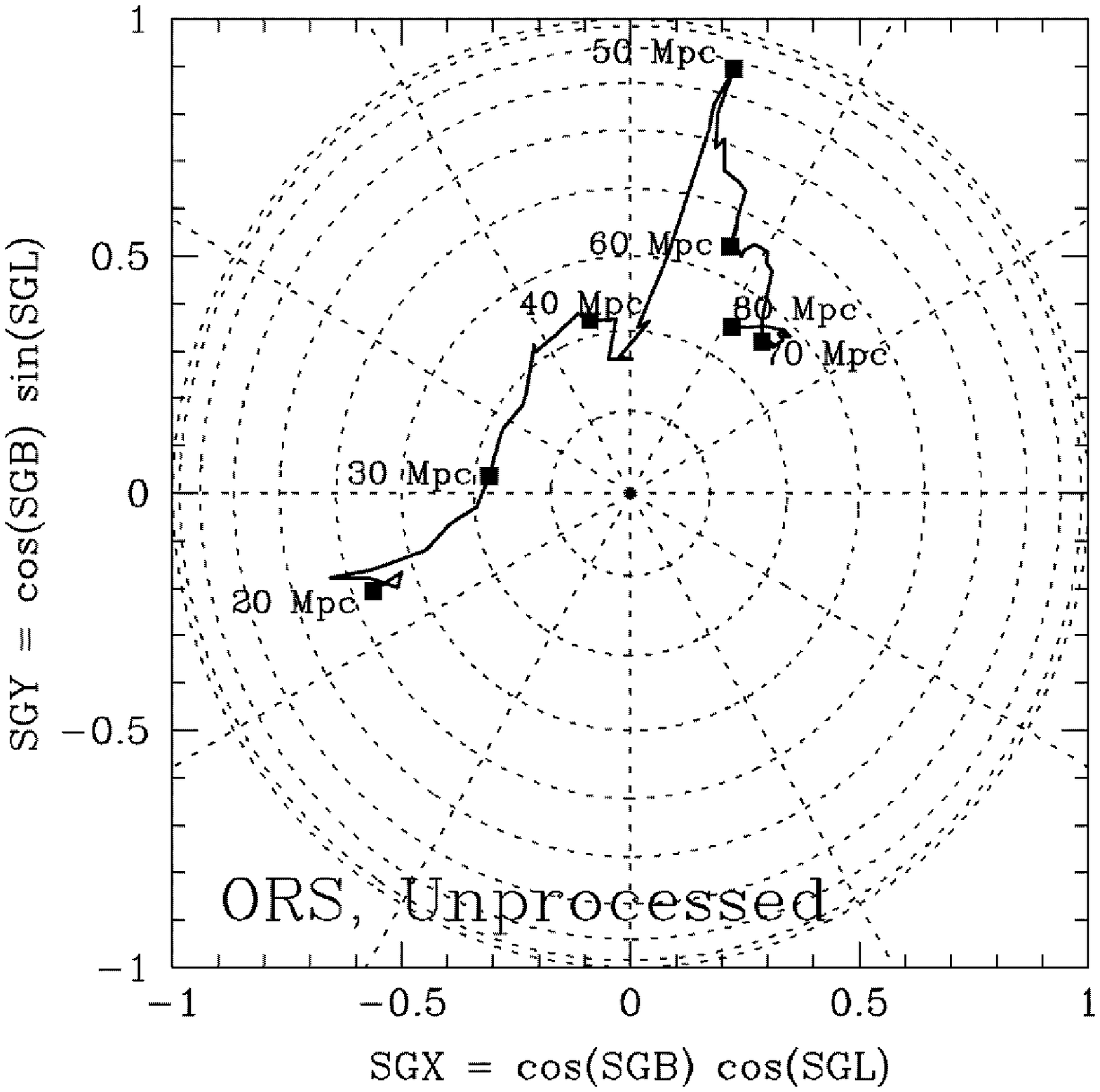,height=7cm,width=7cm,clip=}}
\end{center}
\caption{
   ORS : The direction of the $Z$ axis 
from the PCA analysis (using direct summation over the galaxies)
                             relative to the de Vaucouleurs' pole. 
         Values of $R_{max}$ are indicated. 
}
\end{figure}

\begin{figure}
\begin{center}
\parbox{15cm}{\psfig{figure=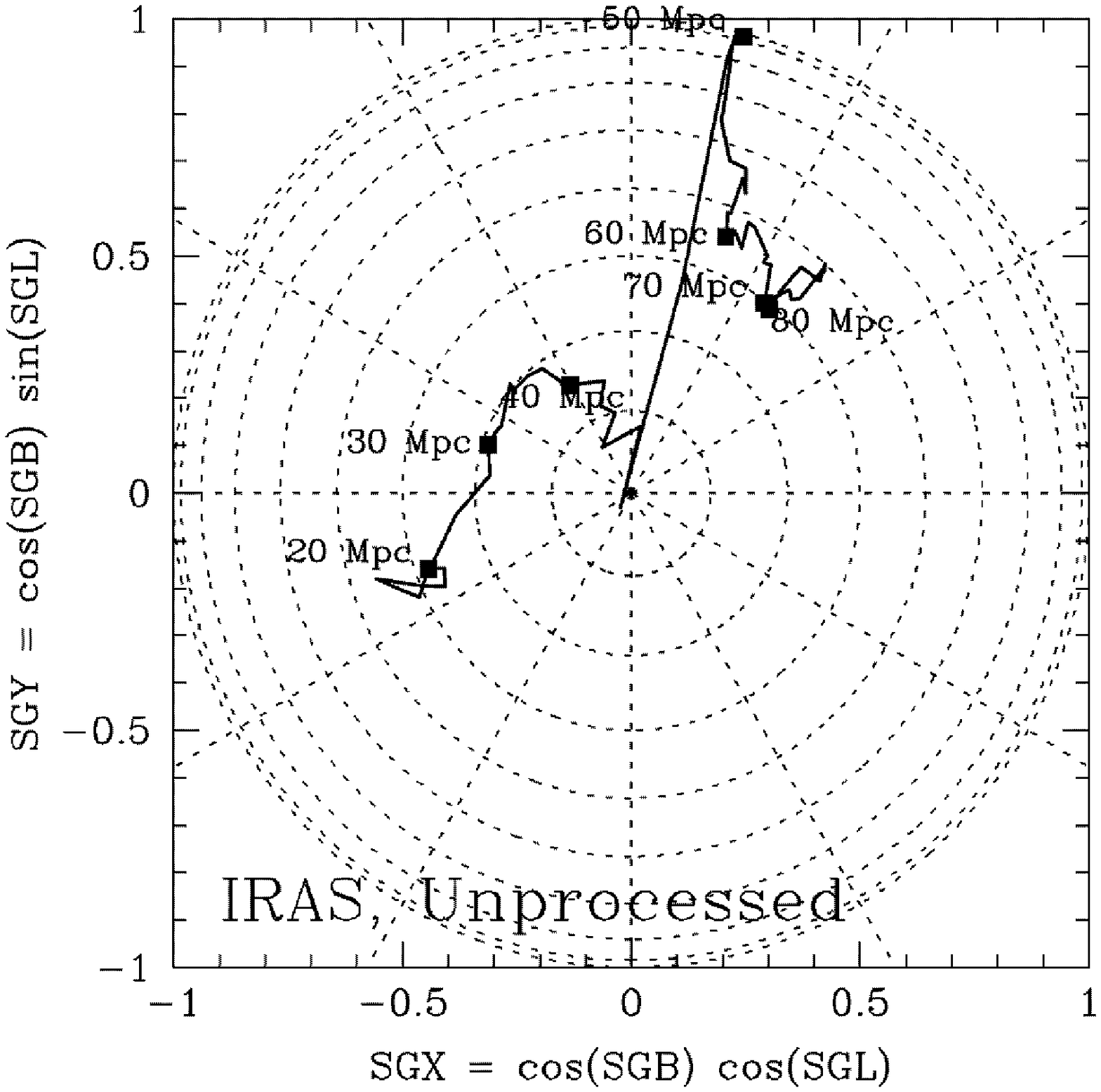,height=7cm,width=7cm,clip=}}
\end{center}
\caption{
   \iras\ : The direction of the $Z$ axis from the PCA analysis
(using direct summation over the galaxies)
                             relative to the de Vaucouleurs' pole. 
         Values of $R_{max}$ are indicated.
}
\end{figure}

\begin{figure}
\begin{center}
\parbox{10cm}{\psfig{figure=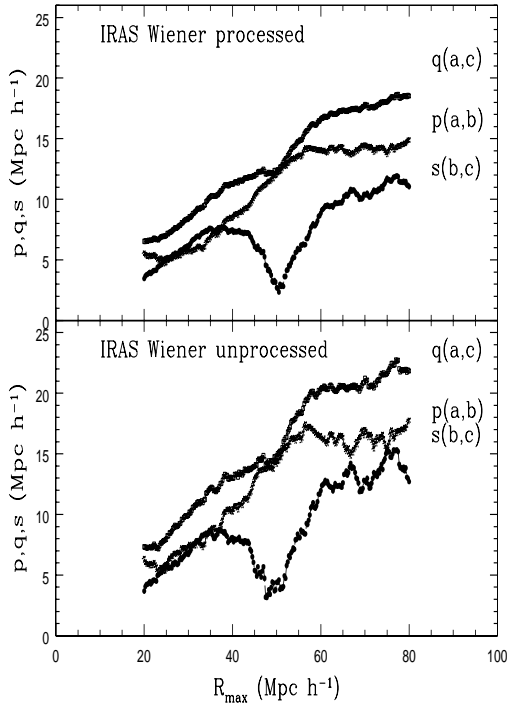,height=10cm,width=7cm,clip=}}
\end{center}
\caption
 {The quantities  $p,q,s$ vs. $R_{max}$  (cumulative) in PCA axes
                         using spherical harmonic reconstruction
                         of \iras\ and analytic MoI.
                        Top: corrected for redshift distortion
                               and noise by Wiener filtering,
                               with parameters 
                            $\beta=0.7, \Gamma=0.2, \sigma_8=0.7$
                               (reconstruction out to 200 \Mpc).
                          Bottom : using raw harmonics. }
\end{figure}

\begin{figure}
\begin{center}
\parbox{15cm}{\psfig{figure=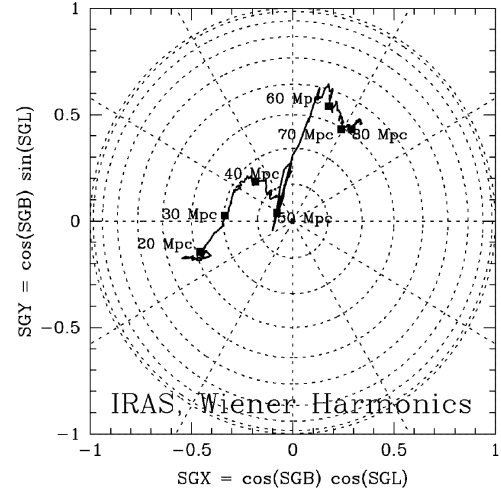,height=7cm,width=7cm,clip=}}
\end{center}
\caption{
   Wiener filtering of \iras\ : The direction of the $Z$ axis (as found by PCA)
                             relative to the de Vaucouleurs' pole. 
         Values of $R_{max}$ are indicated.}
\end{figure}

\begin{table}
\centerline{{\bf Table~1}. SGP parameters from PCA analysis}
\vskip 0.2 true cm
\hrule
\vskip 0.2 true cm
\tabskip=1em plus2em minus.5em
\halign to\hsize
{#\hfil&&\hfil#&\hfil#&\hfil#&\hfil#&\hfil#\cr
$R_{min}$ & $R_{max}$ & $a$  & $b$ & $c$ & 
$\Delta(a^2)$ & $L_z$ & $B_z$  \cr
\noalign{\smallskip\hrule\smallskip
\centerline {ORS}\medskip}
0 & 20 & 9.9  &  8.1  &  7.1   &  2.4  &  160 & 53  \cr
0 & 30 & 14.8 & 12.4  &  11.0  &  5.1  &  173 & 72 \cr
0 & 40 & 19.9 & 16.2  &  14.3  &  7.7  &  104 & 68 \cr
0 & 50 & 25.8 & 20.7  &  20.0  & 12.7  &   76 & 23 \cr
0 & 60 & 31.8 & 26.5  &  22.7  & 17.8  &   67 & 56 \cr
0 & 70 & 35.6 & 31.1  &  27.0  & 24.1  &   48 & 65 \cr
0 & 80 & 39.8 & 35.2  &  31.0  & 33.0  &   58 & 65 \cr
  &    &      &       &        &       &      &    \cr
20& 40 & 21.7 & 17.7  &  15.4  &  9.5  & 101  & 66 \cr 
40& 60 & 35.1 & 29.2  &  24.8  & 23.7  &  65  & 55 \cr
60& 80 & 45.8 & 40.3  &  36.2  & 57.4  &  52  & 68 \cr
\noalign{\bigskip
\centerline {\iras}\smallskip}
0 & 20 & 10.3 &  8.1  &  7.2  &  3.4  &  160  & 62  \cr
0 & 30 & 14.9 &  13.0 & 11.1  &  6.6  &  162  & 71  \cr
0 & 40 & 19.6 &  16.8 & 14.7  & 10.6  &  120  & 75  \cr
0 & 50 & 25.5 &  21.3 & 20.8  & 16.4  &   76  & -6  \cr
0 & 60 & 31.2 &  26.5 & 23.6  & 21.9  &   69  & 55  \cr
0 & 70 & 35.1 &  30.7 & 27.9  & 29.3  &   54  & 60  \cr
0 & 80 & 39.6 &  35.3 & 32.8  & 39.0  &   52  & 61  \cr
  &    &      &       &        &       &      &    \cr
20& 40 & 21.3 & 18.0  &  15.8  & 12.8  & 113  & 72 \cr 
40& 60 & 34.2 & 29.0  &  25.8  & 28.8  &  67  & 53 \cr
60& 80 & 44.9 & 40.5  &  38.3  & 65.9  &  44  & 60 \cr
\noalign{\bigskip
\centerline {\iras\ Wiener}\smallskip}
0 & 20 & 10.1 &  8.5  &  7.8  &  -  &  197  & 61  \cr
0 & 30 & 14.6 & 13.4  & 11.9  &  -  &  176  & 70  \cr
0 & 40 & 19.5 & 17.6  & 15.9  &  -  &  135  & 75  \cr
0 & 50 & 24.5 & 21.4  & 21.2  &  -  &  154  & 85  \cr
0 & 60 & 29.7 & 26.2  & 24.6  &  -  &   72  & 55  \cr
0 & 70 & 34.0 & 30.8  & 29.0  &  -  &   61  & 60  \cr
0 & 80 & 38.3 & 35.3  & 33.5  &  -  &   56  & 59  \cr
}
\smallskip\smallskip\hrule
\medskip

\noindent 
{\bf Comments on Table 1:}
$R_{min}$, $R_{max}$, $a, b$ and $c$ are in $h^{-1}$ Mpc.
The values of $a, b$ and $c$ are for $f_{bg}=0$.
The Poisson error per axis $\Delta (a^2)$ is derived using eqs. (15-17).
No errors are given in the Wiener case, as this is a smoothed field
(although one can calculate the scatter around the mean).

\noindent $L_z$ and $B_z$ (both in degrees) give the direction   
in standard de Vaucouleurs' coordinates of the $Z$ component 
as found by PCA. The angle between this direction and 
de Vaucouleurs' SGP pole is simply $\theta_z = 90^{\circ} - B_z$.
\end{table}

\section{Wiener reconstruction} 

In addition to the problem of shot-noise,
the density field measured from redshift surveys also suffers from
redshift distortions due to peculiar velocities (see Hamilton 1998 for
a review).
One approach which  deals with both problems  is Wiener filtering 
(e.g. Lahav \etal\ 1994, Fisher \etal\ 1995b).
In particular, Fisher \etal\ (1995b) expanded the density field
$\delta({\bf r})$ in terms of 
spherical harmonics and Bessel functions.
By assuming a prior model for the power spectrum, 
the coefficients of the expansion can be corrected for the 
redshift distortion  and for the shot-noise.
This  reconstruction is optimal in the 
minimum variance sense.
In this approach,  the smoothed density field approaches the mean density 
at large distances. This does not mean necessarily that the SGP 
itself disappears at large distances; it merely reflects our ignorance
of the density field where the data are poor. 

This Wiener approach applied to the density field
 $\delta$ also gives the optimal
 reconstruction for any property which is linear in $\delta$.
In particular, if we seek the optimal reconstruction of the
moment of inertia (eq.~7) it can be re-written (for $\bar {\bf x} =0$) as 
$$
{{\tilde C}_{ij} } = 
\left( { 3 \over { 4 \pi R_{max}^3} } \right) I_{ij} +
 \left[ 1 - f_{bg} \right] { R^2 \over 5} 
\delta_{ij}^K
\qquad ,
\eqno (16)
$$
where 
$$ 
I_{ij} = \int_R \delta ({\bf r} ) x_i x_j dV 
\qquad .
\eqno (17)
$$
Webster, Lahav \& Fisher (1997) give analytic expressions
for $I_{ij}$ in terms of the reconstructed
coefficients $\delta_{lmn}^R$, and show that 
only harmonic modes $l=0$ and $l=2$ contribute. 

We apply this technique below to the reconstructed, real-space density
field of \iras\ galaxies, assuming as priors
$\beta\equiv\Omega^{0.6}/b= 0.7$, normalization $\sigma_8 = 0.7$ 
and a CDM power spectrum with
shape parameter $\Gamma =0.2$. 

   Figure 10 shows the $p,q$ and $s$  
   for \iras\ before and after Wiener filtering
   in spherical harmonic presentation.
   Figure 11 shows the variation of the minor PCA axis with $R_{max}$.
   There is good agreement with the results of 
   direct summation shown in Figure 9, indicating that 
   shot noise and redshift distortion do not introduce big systematic effects
   in our statistics. 
   Unfortunately, the Wiener procedure cannot easily be applied to 
   the ORS sample, as it only covers $|b|>20^\circ$.

\section{Interpretation of the MoI results}

The behaviour of the MoI axes (Figures 6-10) 
for both ORS and \iras\ indicate that the SGP
is a structure far more complicated than 
a homogeneous ellipsoid or a slab.
In the local universe
there is `contamination' due to clusters and voids
below and above the SGP. 
To get further insight to the structure that generates the observed
MoI, we generated mock realizations of the \iras\ sample with more
complicated geometries than the simple ellipsoids used in \S 4.4,
trying to mimic the significant overdensities seen in the real universe.  In
particular, we put in: 
(i) an ellipsoid with dimensions of $70\times 70 \times 10 (\Mpc)^3$
and overdensity $\delta=1$ aligned with the Supergalactic plane, 
(ii) a sphere of radius $7.5 \Mpc$ and overdensity 
$\delta=2.5$,
representing a Virgo-like cluster
at $(SGX,SGY,SGZ)= (5,10,-2) \Mpc$,   
(iii) a sphere of radius $11 \Mpc$ and overdensity 
$\delta=3$,
representing a Perseus-Pisces-like supercluster
at  $(SGX,SGY,SGZ)= (49, -23, 0) \Mpc$,   
and
(iv) a sphere of radius $10 \Mpc$ and overdensity 
$\delta=3$,
representing a Great Attractor-like
structure at $(SGX,SGY,SGZ)= (-32, 0, 0) \Mpc$.   

Figure 12 shows $(p,q,s)$  averaged over 100 realizations 
in fixed and PCA coordinates. It is interesting to contrast this Figure
with Figures 6, 7  and 10.
For example we see that the dips in $s$ at $\sim 25 \Mpc$ 
 $50 \Mpc$ in the real data (Figure 6) 
are reproduced here (albeit with not quite the same amplitudes),
especially with fixed axes. 
This toy example illustrates that the observed behaviour of $p, q$,
and $s$ can be
accounted for by adding few major superclusters to 
a pancake-like structure. 
We note that the toy model in Figure 12
does not reproduce the observed 
growth of $p,q$ and $s$ on large scales, which might 
be due to shot noise.

We note that recent simulations for a variety 
of models (e.g., Bond \etal\ 1996, Jenkins \etal\ 1998)
show a web of filaments and sheets, with `knots' at their intersection,
or alternatively,  `arms stretching from clusters'.
It may well be that Perseus-Pisces and the Great Attractor  
represent such knots.
Hence the MoI `shape-finder' statistic, when applied to many points in space,
could be an efficient way of detecting such patterns and discriminating 
between models.

\begin{figure}
\begin{center}
\parbox{10cm}{\psfig{figure=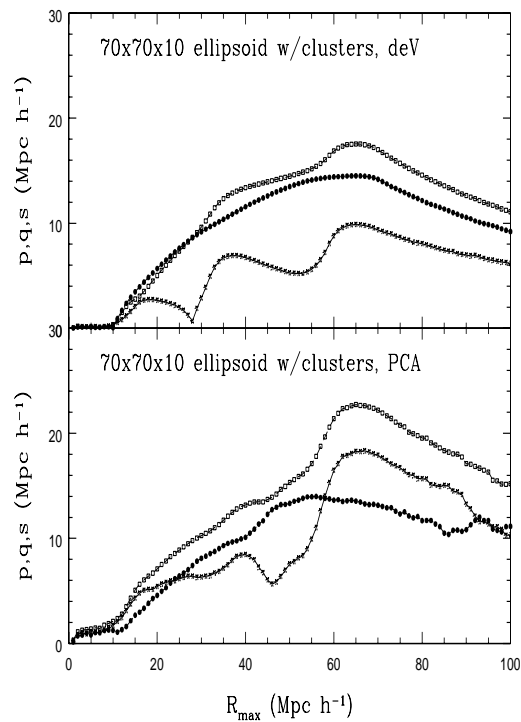,height=10cm,width=7cm,clip=}}
\end{center}
\caption{\label{mock_clusters} 
 The quantities  $p,q,s$ vs. $R_{max}$  (cumulative) in fixed and PCA axes
 for a mock realizations of an oblate ellipsoid plus three clusters,
as described in the text.  The upper panel uses fixed supergalactic
coordinates, while the lower panel uses the axes found by the PCA
analysis. Results were averaged over 100 realizations.}

\end{figure}

\section{Conclusions}

We have presented a Moment of Inertia 
approach to study  
the existence and extent of the planar structure 
in the local galaxy density field, 
the Supergalactic Plane (SGP).

Our main conclusions from the present analysis are:

(i) The SGP is not well-modeled as a homogeneous ellipsoid.

(ii) The Moments of Inertia analysis of the galaxy density field 
shows that 
the normal to the plane is within $\theta_z\sim 30^\circ$ of the standard 
SGP Z-axis,  out to a radius of 80 \Mpc, for both ORS and \iras.
However, the normal changes direction with $R_{max}$, 
in particular at $\sim 50 \Mpc$,
due to the presence of Perseus-Pisces and the Great Attractor.
The shape too varies with $R_{max}$, and in fact
at $50 \Mpc$ the SGP looks more like a dumbbell than 
a pancake.
Based on Figures 6, 7 and 10 
one may argue that the 'plane' terminates at $\sim 40 \Mpc$.
However, we note that the axes  the shell $(60-80) \Mpc$
are aligned with the standard SGP to within $30^\circ$.  

(iii) The density contrast in the slab with   $R_{max} = 40 \Mpc$ and 
thickness of $20 \Mpc$ centered on the Local Group and aligned along
the Supergalactic axes 
is $\delta_{sgp} \sim 0.5$ 
for both  ORS and \iras. 

(iv) The SGP axes and  density contrast 
are similar for both ORS and \iras.

(v) An optimal minimum variance 
reconstruction (Wiener filtering in 
spherical harmonics representation), 
which corrects \iras\ for both  redshift distortion
and shot noise, yields  similar misalignment of angle and axes.

(vi) It may well be that Perseus-Pisces and the Great Attractor  
represent `knots' in a web of filaments and sheets, as
seen in recent $N$-body simulations.
This complicated structure calls to question the notion of 
a single coherent feature on Gpc scales (as claimed e.g. 
by Tully 1987, 1987), and the possible connectivity between the 
Great Attractor and the Shapley Supercluster at 
$\sim 140 \Mpc$ (e.g. Scaramella \etal\ 1992).

Regarding the wider
cosmological implications, the present analysis 
is as an example of applying the MoI at only one point,
centred at us. 
It is of interest to compare the SGP to other structures, 
e.g. the Great Walls seen in the CfA and SSRS surveys.
One can  calculate moments of
inertia at many points in space and  contrast the statistical
behaviour of the axes with $N$-body simulations.
Results 
from  volume limited subsets of $N$-body simulations 
show indeed that the MoI 
statistic quantifies  e.g. differences in filamentary structure between 
Cold and Hot dark matter models (e.g. Dave \etal\ 1997, Webster 1998). 
Unfortunately, the flux limited ORS and \iras\ catalogues are too shallow to 
apply the statistic in other points in space.
Volume limited
subsets from new large redshift surveys (e.g. 2dF and SDSS) would be
suitable for such shape-finding algorithms.

It also is possible to improve the above analysis by calculating the moment
of inertia for a pre-selected region above a certain
density-threshold.  
It would also be interesting to repeat the analysis for 
different morphological types
and weighting schemes; for example, by calculating
luminosity-weighted, rather than number weighted galaxy counts.  On
the statistical side, other more general shape-finders can be applied,
e.g. by utilizing the Minkowski functions (e.g. Mecke, Buchert \&
Wagner 1994; Sahni, Sathyaprakash \& Shandarin 1998).

\bigskip
\bigskip

{\bf Acknowledgments:} 
We thank J. Bagla, A. Dekel, K. Fisher, D. Lynden-Bell,  
P. Monaco and S. Shandarin  for helpful discussions and comments.
MD  acknowledges support from NSF grant AST95-28340.
JPH was supported by the
Smithsonian Institution and NASA NAGW-201. 
OL thanks the Weizmann Institute (Israel), 
the Ango-Australian Observatory and ATNF/CSIRO (Australia) 
for their hospitality. 
MAS acknowledges support from the Alfred P. Sloan Foundation, Research
Corporation, and NSF grant AST96-16901.  
AMW acknowledges the receipt of a PPARC studentship.

\end{document}